\documentclass{article} 
\usepackage{iclr2020_conference,times}

\usepackage{amsmath,amsfonts,bm}

\def\eqref#1{equation~\ref{#1}}

\def\1{\bm{1}}

\DeclareMathAlphabet{\mathsfit}{\encodingdefault}{\sfdefault}{m}{sl}
\SetMathAlphabet{\mathsfit}{bold}{\encodingdefault}{\sfdefault}{bx}{n}

\usepackage{hyperref}
\usepackage{url}

\title{Artificial Intelligence for Global Health: learning from a decade of digital transformation in health care   
}

\author{Varoon Mathur \\
The AI Now Institute\\
New York University\\
New York City, NY 10013, USA \\
\texttt{varoon@ainowinstitute.org} \\
\And
Saptarshi Purkayastha, PhD \\
School of Informatics and Computing \\
Indiana University-Purdue University Indianapolis\\
Indianapolis, IN 46202, USA \\
\texttt{saptpurk@iupui.edu} \\
\AND
Judy Wawira Gichoya, MD, MS \\
Department of Radiology and Imaging Sciences \\
Emory University \\
Atlanta, GA 30322, USA \\
\texttt{judywawira@emory.edu}
}

\iclrfinalcopy 
\begin{document}

\maketitle

\begin{abstract}
The health needs of those living in resource-limited settings are a vastly overlooked and understudied area in the intersection of machine learning (ML) and health care. While the use of ML in health care is more recently popularized over the last few years from the advancement of deep learning, low-and-middle income countries (LMICs) have already been undergoing a digital transformation of their own in health care over the last decade, leapfrogging milestones due to the adoption of mobile health (mHealth). With the introduction of new technologies, it is common to start afresh with a top-down approach, and implement these technologies in isolation, leading to lack of use and a waste of resources. In this paper, we outline the necessary considerations both from the perspective of current gaps in research, as well as from the lived experiences of health care professionals in resource-limited settings. We also outline briefly several key components of successful implementation and deployment of technologies within health systems in LMICs, including technical and cultural considerations in the development process relevant to the building of machine learning solutions. We then draw on these experiences to address where key opportunities for impact exist in resource-limited settings, and where AI/ML can provide the most benefit.
\end{abstract}

\section{Introduction}

Achieving the 2030 Agenda for Sustainable Development Goals (SDGs), especially ones that focus on the promotion of health and well-being through the reduction of preventable deaths, the spread of epidemics, and universal access to care, are especially difficult in LMICs due to a lack of resources, capacity, infrastructure, and affordable medicines and treatment \citep{vinuesa_role_2020}. Here, machine learning and related AI technologies have been broadly identified as a potential tool for addressing the unmet health needs for vulnerable and marginalized populations in LMICs \citep{wahl_artificial_2018}. 

While early research success has shown promise of clinical relevance of these systems, there exists still a great number of challenges for the meaningful development and deployment of AI in the field of health care. A lack of prospective and clinical validation studies to understand the added benefit for patients is still lacking \citep{topol_high-performance_2019}, and the privacy and security of individually identifiable health data that is used for training and testing of such algorithms are still unresolved concerns in the wake of anonymized data that can be readily identified \citep{hejblum_probabilistic_2019}. 

\subsection{Complexities in Capacity in the LMIC Context}

However, the LMIC context adds additional layers of complexity in regards to the integration of AI/ML solutions within health care.  For example, a large proportion of deep learning and computer vision research is motivated by the need to address the “shortage of experts” who can interpret medical imaging \citep{rajpurkar_chexnet_2017} or that can “shorten the gap between doctors at different levels” or between hospitals \citep{rajpurkar_chexnet_2017}, \citep{li_artificial_2019}. These frameworks necessarily lead to experiments that focus primarily on reporting metrics such as area under the curve (AUC) in order to measure accuracy and predictive validity, which may not necessarily be indicative of clinical outcomes \citep{topol_high-performance_2019}. Yet, such narratives of clinicians being “beat” by AI-systems continue to dominate mainstream headlines \citep{hateley_doctor_2017}. These narratives have been pushed back against by current scholarship, as \cite{topol_high-performance_2019} lays out, given that the goal instead should be of “synergy” between the clinician and the AI system. More recent work has shown that these “human-machine partnerships” perform much better than either agent on its own, as demonstrated by a recent study recruiting 13 radiologists to work together with an AI-mediated platform for chest radiograph diagnosis \citep{patel_humanmachine_2019}.

Where does this leave resource-poor health systems? On the one hand, machine learning aimed at replacing clinicians or to fill gaps where expertise lacks does not align with current demographic trends within LMICs, and in fact aligns more with high income countries instead. More than 40 countries experience shrinking working-age populations, defined as ages 15-64 years of age, while nearly 80 of counties within the United States specifically lost working-age adults between 2007 and 2017, with further losses projected in the coming years \citep{hart_shrinking_2019}. Meanwhile, African nations are collectively projected to eclipse the workforce of Asia by the end of the 21st century, with a shift towards youth ready for employment \citep{tanzi_africas_2019}. The use of ML to therefore compensate for a lack of expertise, seems to have little justification in the context of a growing workforce that can indeed fill those gaps themselves. 

Conversely, machine learning tools and algorithms to be used in tandem with expert physicians does not grapple with the general resource constraints present in LMICs. Not only is it costly and infeasible to recruit multiple experts to work in alongside AI systems, but the infrastructure of AI systems themselves can become quite expensive to procure \citep{bresnick_artificial_2018}. A study conducted to assess the cost of validation and integration of a chronic kidney disease analytics application, for example, was estimated to cost nearly \$220,000 for a single hospital system in the United States \citep{sendak_barriers_2017}. This necessarily prices out health systems in the developing world from implementing these costly systems. Moreover, pricing models that are distributed as partnerships where a hospital gets access to the algorithms in exchange for the provision of labeled data do not apply to LMICs. Patients crossover hospitals based on where they get sick, and the distribution of specialist hospitals would mean a partnership with such hospital is inaccessible to people from all over the country who sort service at a specific point in time. These specialist hospitals remain in the large cities, hence such incentives can exacerbate access to equitable healthcare.

\subsection{Further Considerations from an LMIC Perspective}

Early implementations of digital health technologies traditionally involved a hybrid of paper, mobile and computer based tools. This is because patients kept a complete record of their health system, and these records would not be available across institutions. Moreover, poor infrastructure meant that backups could fail (before adoption of cloud technologies), and electricity could be unavailable for weeks. Therefore, redundancies became critical in order to minimize interruptions in patient care. In one specific case, at one of the institutions in Western Kenya, paper and electronic records were duplicated for years, resulting in a large cost of paper and increased burden to providers who would need to deal with duplicate data entries. This would inherently result in incomplete records, specifically due to the isolated and disruptive nature of the EMR/mHealth implementation. A further example of this, a given female patient with HIV would be enrolled in the EMR system, and after pregnancy, would be referred to the antenatal clinic, but after delivery would resume care in the corresponding Adult HIV clinic while her child would be referred to the another clinic. With fragmented implementations of health systems, data simply could not be captured by both clinics. It is precisely these challenges in such workflows that exist given resource-limited contexts, and it is exactly the challengers that ML developers must be cognizant of, in regards to where gaps may arise. 

Further, mobile devices are used for many daily activities including healthcare and finance, and early systems explored web applications, SMS and Interactive voice response systems (IVR). Their success has historically been dependent on cost. IVR systems would require memorization of multiple codes to get to the desired options, while the cost of calls and SMS would be prohibitive. Today, there are texting applications like WhatsApp that are used for communication, especially in groups. These technologies will more than likely be incorporated into the ML based solutions given, however these platforms will inherently have different, and often unaligned, incentives. When Facebook groups were encouraged for use for the purposes of health education with countries such as Bhutan, where Facebook would be used to establish a virtual consulting and referral system, misinformation would not have been a primary concern. However, this is an example that today would be much more difficult in the era of misinformation, that has for example bolstered the voices of anti-vaxxers. We posit the notion that delivering health results through these platforms can be harmful, since incentives do not always align in the delivery of health information. This is exemplified by studying the patterns of mobile ownership, whereby the head of the household can own the phone, and hence when these phones are used for patient care or delivery of critical information then differential privacy is inadequate since privacy is more than a technical challenge.

Finally, lack of consensus on standards and guidelines presents a conundrum when developing ML/AI tools for global health. For example, caring for HIV patients in mid-1990's and early 2000s, at the peak of the HIV/AIDS crisis, gave way to the criteria for care providers initiating antiretroviral (ARV) treatment would be a white-blood cell count with a minimum of 200 cells. The reasoning at the time was inextricably linked to limited resources, and hence a decision was made to treat the sickest patients as a priority. As funding and an abundance in resources increased, the minimum count criteria was raised to 350, with a clear directive to initiate ARV treatment as early as possible. This contextual information is not frequently captured, and as far as the development of ML tools for LMICs is concerned, can cause harm when such guidelines are implemented and deployed at scale. This means that the LMICs are also not uniform, therefore requiring the localization of theses technologies in order to confer power to the communities it intends to serve.

Despite these challenges, several examples of successful and important technology-driven solutions deployed in resource-limited health care systems exist that can serve as a practical guide for the development of ML solutions in LMICs. We highlight and provide a broad analysis of such systems to draw out potential best practices, that can inform critical targets in the global health landscape, and that can provide opportunities for machine learning to have meaningful impact for patients in developing countries.

\section{Case Analysis of Successful Implementations}

While both \cite{wahl_artificial_2018} and \cite{vinuesa_role_2020} highlight examples of ML systems that promise potential benefit to patient health in LMICs specifically within the context of the Millennium Development Goals (MDGs) and SDG 3, none provide critical insight into how these models are integrated, as all are model validation studies. We instead focus on two non-AI examples that capture key lessons from their development and deployment, as well as their process of integration in hospital and health care systems within African nations. 

\subsection{OpenMRS}

The OpenMRS (open source medical record system) software was originally designed with the need to address the unmet need of HIV patients in mind, but soon adopted a broader vision of creating a robust and scalable open source system for health care delivery \citep{sims_write_2019}. Data management solutions that could offload paper documentation for hospitals, not only in the wake of a staggering number of individuals living with HIV/AIDS, but also given those living with multi-drug resistant tuberculosis, were critical for the uninterrupted operations of these health systems. Over a period of more than a decade, OpenMRS has seen its adoption in hundreds of health facilities spanning 42 countries including South Africa, Kenya, Rwanda and Lesotho, among others \citep{seebregts_openmrs_2009}. Its success not only attributable to its considerably reduced cost, but to the robust open source community (the OpenMRS Implementers Network) that could be depended on to provide technical support, analogous to what would normally be provided by a commercial entity \citep{seebregts_openmrs_2009}. 

Several key factors have been identified as part of the success and growth of OpenMRS, and as \cite{sims_write_2019} describes, the dedication and commitment of involved project members at the outset of software construction were critical in developing a robust open source community that soon followed. Being “medical professionals first and technologists second” not only allowed for the lived experiences of those caring for patients in these specific communities to shape the development process from the beginning, it strengthened a developer community to come together rapidly given their similar backgrounds, predicated on the belief that writing code really was “saving lives” \citep{sims_write_2019}. This rapid growth in membership also aided directly in bringing other stakeholders such as intergovernmental organizations to the table to spur new areas of deployment \citep{seebregts_openmrs_2009}.

Of note - that such a large proportion of those contributing to the open source software were medical professionals and clinicians in some cases, meant that a challenge that presented itself during the development of OpenMRS was that a gap existed between available personnel and those who were adequately trained. The average time to train a developer was found to be close to 24 months, at which it was likely they were already being recruited for higher paying developer roles. Therefore, successful EMR and mHealth tools evolved to rely on an architecture that was simplified using a web application (initial attempts of OpenMRS were primarily Java based) with a more stable back end. This provides a critical lesson for the use of ML tools in similar contexts, as it will be more than likely that personnel training needs are not just for developers, but for future health care providers who will likely engage daily with these systems. This is also further evidence that shifting work to a ML system that operates autonomously without creating adequate integration in the workflow of health care workers, may only have limited short term gains for diagnosis but experience an overall reduction in capacity building.

\subsection{DHIS2}

Another open source web-based implementation that has seen marked success in LMICs over a sustained period of more than two decades has been the District Health Information System (DHIS2), designed as a health management information system to store and retrieve medical data, but also census and community data, so as to improve the decision making processes for any given health system \citep{karuri_dhis2_2014}. The history of DHIS2 itself is rooted deeply in a vision to combat health inequity, as the development was originally started by the Health Information System Programme (HISP) within South Africa, in an attempt to address the disparities in health care due to apartheid \citep{braa_dhis2_2017}.  The formal partnership between HISP and the researchers from the University of Oslo brought together seemingly contrasting worlds, that of anti-apartheid activists and public health officials, alongside Scandanavian programmers with a rich history of open source development \citep{karuri_dhis2_2014}. DHIS2 currently is utilized 67 different countries and countless more health systems, especially South Africa, Ghana, Uganda, and Rwanda \citep{braa_dhis2_2017}.

Similarly to OpenMRS, DHIS2 early success could be attributed to the strong open source community forged through projects that connected developers and communities from a wide-range of stakeholder groups such as health regulatory agencies, NGOs, and academics early on in the development process \citep{karuri_dhis2_2014}. In addition, the educational component of HISP projects, specifically in South Africa, not only allowed for the training of “thousands” of health workers to be educated in the development and maintenance of the system, but also allowed for the project to spread to other countries and hospital systems as well \citep{braa_dhis2_2017}. 

\section{Practical Recommendations for AI Health Solutions in LMICs}

Previous reviews of challenges and key lessons for ML deployment in resource-limited settings have heavily focused on data collection, data labeling, and other aspects of the machine learning pipeline that allows for better model training and deployment \citep{SSS101125}. These are of course necessary, and the current bias mitigation strategies of understand the context and use of a data set prior to the training and validation phase are of critical importance. 

However, learning from critical open source software that serves as a backbone for health system infrastructure such as OpeMRS and DHIS2 provides a deeper understanding into the integration that relies much less on the specific algorithms and computational tools, and much more on design and implementation. This is in fact a closely related, but separate issue to clinical validation of machine learning models, as this phase happens after the design phase but not before.

In both cases, the development of software tools incorporated the eventual users of the system as co-designers in a collaborative open source environment, which could effectively be considered a participatory design approach \citep{braa_participatory_2012}. Not only does an open source environment allow for significant cost reduction in hospital procurement \citep{sims_write_2019}, and build trust with the communities they serve, but investing in educational tools and investing in training for care providers to also become code contributors also addressed capacity building among health care workers and administrators. 

This shifts the critical targets for AI/ML in LMICs for health away from stand-alone systems, such as perhaps computer vision systems competing against clinicians, and rather towards health information infrastructure and technology that focuses on capacity building and empowerment of health care workers already on the ground. 

\bibliography{iclr2020_conference}

\begin{thebibliography}{18}
\providecommand{\natexlab}[1]{#1}
\providecommand{\url}[1]{\texttt{#1}}
\expandafter\ifx\csname urlstyle\endcsname\relax
  \providecommand{\doi}[1]{doi: #1}\else
  \providecommand{\doi}{doi: \begingroup \urlstyle{rm}\Url}\fi

\bibitem[Braa \& Sahay(2012)Braa and Sahay]{braa_participatory_2012}
Jørn Braa and Sundeep Sahay.
\newblock Participatory design within the {HISP} network: {Looking} back and
  looking forward.
\newblock pp.\  p 235--256. January 2012.

\bibitem[Braa \& Sahay(2017)Braa and Sahay]{braa_dhis2_2017}
Jørn Braa and Sundeep Sahay.
\newblock The {DHIS2} {Open} {Source} {Software} {Platform}: {Evolution} {Over}
  {Time} and {Space}.
\newblock April 2017.
\newblock ISBN 978-0-262-33812-7.

\bibitem[Bresnick(2018)]{bresnick_artificial_2018}
Jennifer Bresnick.
\newblock Artificial {Intelligence} in {Healthcare} {Spending} to {Hit}
  \${36B}, December 2018.
\newblock URL
  \url{https://healthitanalytics.com/news/artificial-intelligence-in-healthcare-spending-to-hit-36b}.

\bibitem[Hart(2019)]{hart_shrinking_2019}
Kim Hart.
\newblock The shrinking working-age population, July 2019.
\newblock URL
  \url{https://www.axios.com/disappearance-working-age-americans-9cfd00eb-6bbd-465c-8d63-11832eac25c4.html}.

\bibitem[Hateley(2017)]{hateley_doctor_2017}
Peter Hateley.
\newblock Doctor {AI} will see you now.
\newblock \emph{BMJ}, 356, January 2017.
\newblock ISSN 1756-1833.
\newblock \doi{10.1136/sbmj.i6528}.
\newblock URL \url{https://www.bmj.com/content/356/sbmj.i6528}.

\bibitem[Hejblum et~al.(2019)Hejblum, Weber, Liao, Palmer, Churchill, Shadick,
  Szolovits, Murphy, Kohane, and Cai]{hejblum_probabilistic_2019}
Boris~P. Hejblum, Griffin~M. Weber, Katherine~P. Liao, Nathan~P. Palmer,
  Susanne Churchill, Nancy~A. Shadick, Peter Szolovits, Shawn~N. Murphy,
  Isaac~S. Kohane, and Tianxi Cai.
\newblock Probabilistic record linkage of de-identified research datasets with
  discrepancies using diagnosis codes.
\newblock \emph{Scientific Data}, 6\penalty0 (1):\penalty0 1--11, January 2019.
\newblock ISSN 2052-4463.
\newblock \doi{10.1038/sdata.2018.298}.
\newblock URL \url{https://www.nature.com/articles/sdata2018298}.

\bibitem[Karuri et~al.(2014)Karuri, Waiganjo, Orwa, and
  Manya]{karuri_dhis2_2014}
Josephine Karuri, Peter Waiganjo, Daniel Orwa, and Ayub Manya.
\newblock {DHIS2}: {The} {Tool} to {Improve} {Health} {Data} {Demand} and {Use}
  in {Kenya}.
\newblock \emph{Journal of Health Informatics in Developing Countries},
  8\penalty0 (1), March 2014.
\newblock ISSN 1178-4407.
\newblock URL \url{http://jhidc.org/index.php/jhidc/article/view/113}.

\bibitem[Li et~al.(2019)Li, Shen, Xue, Shen, Jing, Wang, Xu, Meng, Yu, and
  Cui]{li_artificial_2019}
Cheng-Xu Li, Chang-Bing Shen, Ke~Xue, Xue Shen, Yan Jing, Zi-Yi Wang, Feng Xu,
  Ru-Song Meng, Jian-Bin Yu, and Yong Cui.
\newblock Artificial intelligence in dermatology: past, present, and future.
\newblock \emph{Chinese Medical Journal}, 132\penalty0 (17):\penalty0
  2017--2020, September 2019.
\newblock ISSN 0366-6999.
\newblock \doi{10.1097/CM9.0000000000000372}.
\newblock URL
  \url{https://journals.lww.com/cmj/fulltext/2019/09050/artificial_intelligence_in_dermatology__past,.1.aspx}.

\bibitem[Patel et~al.(2019)Patel, Rosenberg, Willcox, Baltaxe, Lyons, Irvin,
  Rajpurkar, Amrhein, Gupta, Halabi, Langlotz, Lo, Mammarappallil, Mariano,
  Riley, Seekins, Shen, Zucker, and Lungren]{patel_humanmachine_2019}
Bhavik~N. Patel, Louis Rosenberg, Gregg Willcox, David Baltaxe, Mimi Lyons,
  Jeremy Irvin, Pranav Rajpurkar, Timothy Amrhein, Rajan Gupta, Safwan Halabi,
  Curtis Langlotz, Edward Lo, Joseph Mammarappallil, A.~J. Mariano, Geoffrey
  Riley, Jayne Seekins, Luyao Shen, Evan Zucker, and Matthew~P. Lungren.
\newblock Human–machine partnership with artificial intelligence for chest
  radiograph diagnosis.
\newblock \emph{npj Digital Medicine}, 2\penalty0 (1):\penalty0 1--10, November
  2019.
\newblock ISSN 2398-6352.
\newblock \doi{10.1038/s41746-019-0189-7}.
\newblock URL \url{https://www.nature.com/articles/s41746-019-0189-7}.

\bibitem[Rajpurkar et~al.(2017)Rajpurkar, Irvin, Zhu, Yang, Mehta, Duan, Ding,
  Bagul, Langlotz, Shpanskaya, Lungren, and Ng]{rajpurkar_chexnet_2017}
Pranav Rajpurkar, Jeremy Irvin, Kaylie Zhu, Brandon Yang, Hershel Mehta, Tony
  Duan, Daisy Ding, Aarti Bagul, Curtis Langlotz, Katie Shpanskaya, Matthew~P.
  Lungren, and Andrew~Y. Ng.
\newblock {CheXNet}: {Radiologist}-{Level} {Pneumonia} {Detection} on {Chest}
  {X}-{Rays} with {Deep} {Learning}.
\newblock \emph{arXiv:1711.05225 [cs, stat]}, December 2017.
\newblock URL \url{http://arxiv.org/abs/1711.05225}.
\newblock arXiv: 1711.05225.

\bibitem[Seebregts et~al.(2009)Seebregts, Mamlin, Biondich, Fraser, Wolfe,
  Jazayeri, Allen, Miranda, Baker, Musinguzi, Kayiwa, Fourie, Lesh, Kanter,
  Yiannoutsos, and Bailey]{seebregts_openmrs_2009}
Christopher~J. Seebregts, Burke~W. Mamlin, Paul~G. Biondich, Hamish S.~F.
  Fraser, Benjamin~A. Wolfe, Darius Jazayeri, Christian Allen, Justin Miranda,
  Elaine Baker, Nicholas Musinguzi, Daniel Kayiwa, Carl Fourie, Neal Lesh,
  Andrew Kanter, Constantin~T. Yiannoutsos, and Christopher Bailey.
\newblock The {OpenMRS} {Implementers} {Network}.
\newblock \emph{International Journal of Medical Informatics}, 78\penalty0
  (11):\penalty0 711--720, November 2009.
\newblock ISSN 1386-5056.
\newblock \doi{10.1016/j.ijmedinf.2008.09.005}.
\newblock URL
  \url{http://www.sciencedirect.com/science/article/pii/S1386505608001652}.

\bibitem[Sendak et~al.(2017)Sendak, Balu, and Schulman]{sendak_barriers_2017}
Mark~P. Sendak, Suresh Balu, and Kevin~A. Schulman.
\newblock Barriers to {Achieving} {Economies} of {Scale} in {Analysis} of {EHR}
  {Data}.
\newblock \emph{Appl Clin Inform}, 08\penalty0 (3):\penalty0 826--831, July
  2017.
\newblock ISSN 1869-0327.
\newblock \doi{10.4338/ACI-2017-03-CR-0046}.
\newblock URL
  \url{http://www.thieme-connect.de/DOI/DOI?10.4338/ACI-2017-03-CR-0046}.

\bibitem[Sims et~al.(2019)Sims, Gichoya, Bhardwaj, and Bogers]{sims_write_2019}
Jonathan Sims, Judy Gichoya, Gaurab Bhardwaj, and Marcel Bogers.
\newblock Write {Code}, {Save} {Lives}: {How} a {Community} {Uses} {Open}
  {Innovation} to {Address} a {Societal} {Challenge}.
\newblock \emph{R\&D Management}, 49\penalty0 (3):\penalty0 369--382, June
  2019.
\newblock ISSN 0033-6807, 1467-9310.
\newblock \doi{10.1111/radm.12338}.
\newblock URL \url{https://onlinelibrary.wiley.com/doi/abs/10.1111/radm.12338}.

\bibitem[Tanzi \& Lu(2019)Tanzi and Lu]{tanzi_africas_2019}
Alexandre Tanzi and Wei Lu.
\newblock Africa's {Working}-{Age} {Population} to {Top} {Asia}'s by 2100.
\newblock \emph{Bloomberg.com}, July 2019.
\newblock URL
  \url{https://www.bloomberg.com/news/articles/2019-07-20/africa-s-working-age-population-to-surpass-china-s-by-2100}.

\bibitem[Topol(2019)]{topol_high-performance_2019}
Eric~J. Topol.
\newblock High-performance medicine: the convergence of human and artificial
  intelligence.
\newblock \emph{Nature Medicine}, 25\penalty0 (1):\penalty0 44--56, January
  2019.
\newblock ISSN 1546-170X.
\newblock \doi{10.1038/s41591-018-0300-7}.
\newblock URL \url{https://www.nature.com/articles/s41591-018-0300-7}.

\bibitem[Vinuesa et~al.(2020)Vinuesa, Azizpour, Leite, Balaam, Dignum, Domisch,
  Felländer, Langhans, Tegmark, and Fuso~Nerini]{vinuesa_role_2020}
Ricardo Vinuesa, Hossein Azizpour, Iolanda Leite, Madeline Balaam, Virginia
  Dignum, Sami Domisch, Anna Felländer, Simone~Daniela Langhans, Max Tegmark,
  and Francesco Fuso~Nerini.
\newblock The role of artificial intelligence in achieving the {Sustainable}
  {Development} {Goals}.
\newblock \emph{Nature Communications}, 11\penalty0 (1):\penalty0 1--10,
  January 2020.
\newblock ISSN 2041-1723.
\newblock \doi{10.1038/s41467-019-14108-y}.
\newblock URL \url{https://www.nature.com/articles/s41467-019-14108-y}.

\bibitem[Wahl et~al.(2018)Wahl, Cossy-Gantner, Germann, and
  Schwalbe]{wahl_artificial_2018}
Brian Wahl, Aline Cossy-Gantner, Stefan Germann, and Nina~R. Schwalbe.
\newblock Artificial intelligence ({AI}) and global health: how can {AI}
  contribute to health in resource-poor settings?
\newblock \emph{BMJ Global Health}, 3\penalty0 (4), August 2018.
\newblock ISSN 2059-7908.
\newblock \doi{10.1136/bmjgh-2018-000798}.
\newblock URL \url{https://gh.bmj.com/content/3/4/e000798}.

\bibitem[Weber \& Toyama(2010)Weber and Toyama]{SSS101125}
Julie Weber and Kentaro Toyama.
\newblock Remembering the past for meaningful ai-d, 2010.
\newblock URL
  \url{https://www.aaai.org/ocs/index.php/SSS/SSS10/paper/view/1125}.

\end{thebibliography}
\bibliographystyle{iclr2020_conference}

\end{document}